# Characterization of Saturn's bow shock: Magnetic field observations of quasi-perpendicular shocks

A. H. Sulaiman,[1,2] A. Masters,[1] M. K. Dougherty[1]

Corresponding author: A.H. Sulaiman (ali-sulaiman@uiowa.edu)

[1]Space and Atmospheric Physics, Blackett Laboratory, Imperial College London, London, UK.

[2]Department of Physics and Astronomy, University of Iowa, Iowa City, Iowa, USA.

**Abstract**

Collisionless shocks vary drastically from terrestrial to astrophysical regimes resulting in radically different characteristics. This poses two complexities. Firstly, separating the influences of these parameters on physical mechanisms such as energy dissipation. Secondly, correlating observations of shock waves over a wide range of each parameter, enough to span across different regimes. Investigating the latter has been restricted since the majority of studies on shocks at exotic regimes (such as supernova remnants) have been achieved either remotely or via simulations, but rarely by means of in-situ observations. Here we present the parameter space of $M_A$ bow shock crossings from 2004-2014 as observed by the Cassini spacecraft. We find that Saturn's bow shock exhibits characteristics akin to both terrestrial and astrophysical regimes ($M_A$ of order 100), which is principally controlled by the upstream magnetic field strength. Moreover, we determined the $\theta_{Bn}$ of each crossing to show that Saturn's (dayside) bow shock is predominantly quasi-perpendicular by virtue of the Parker spiral at 10 AU. Our results suggest a strong dependence on $M_A$ in controlling the onset of physical mechanisms in collisionless shocks, particularly non-time stationarity and variability. We anticipate our comprehensive assessment will yield deeper insight into high $M_A$ collisionless shocks and provide a broader scope for understanding the structures and mechanisms of collisionless shocks.

**Introduction**

The magnetosphere of Saturn, like those of all magnetised planets in the solar system, is a largely impenetrable blunt obstacle to the super-fast magnetosonic solar wind flow. In other words, the continuous stream of plasma is travelling from the Sun at a relative speed greater than that at which the obstacle's presence can be propagated through the plasma fluid [*Burgess*, 1995]. As a result, a detached fast shock wave is formed upstream of the obstacle slowing down and thermalizing the flow particles and thus conserving mass, momentum, and energy. The collisional scale of the solar wind is understood to be many orders of magnitude greater than the length of the shock's transition layer and thus coupling between the particles and electromagnetic fields come into effect to regulate the flow states such that the conservation laws are maintained across the shock [*Burgess and Scholer*, 2015]. Such interactions complexity relates to a dependence on various control parameters influencing the shock structure especially when separating their effects since two or more parameters can generally change together. These parameters include Mach numbers $M$ (ratio of the relative speed to a characteristic wave speed namely fast magnetosonic, Alfvén, or sound speeds), plasma $\beta$ (ratio of thermal to magnetic pressures) and magnetic contact angle $\theta_{Bn}$ (angle between the upstream magnetic field vector and local shock normal $\hat{\boldsymbol{n}}$). This quantity is used to distinguish between two categories of shocks: quasi-parallel and quasi-perpendicular where $0° \leq \theta_{Bn} < 45°$ and $45° \leq \theta_{Bn} < 90°$ respectively.

*Criticality*

Consider the formation of a fast magnetosonic shock ($M_f > 1$) which is most commonly found in planetary environments. At low Mach numbers just above unity, the shock could be capable of dissipating the required energy entirely through resistivity from "anomalous collisions" owing to the collective interactions between particles and fields [*Kennel and Sagdeev*, 1967]. As the arbitrary flow speed continues to increase (or equivalently a change in the thermodynamic properties such as temperature, density, and pressure decrease the characteristic speeds), the Mach number continues to rise accordingly. Inevitably a point will be reached where the shock is incapable of converting an excess of ram energy upstream into thermal energy downstream. This character of the shock, in maintaining the required "heating" solely by anomalous resistivity, is referred to as its criticality. The largest critical Mach number is $M_c \approx 2.8$, depending on $\theta_{Bn}$ and $\beta$ [*Marshall*, 1955]. Shocks with $M < M_c$ are subcritical and $M > M_c$ are supercritical; the latter being the most commonly observed in planetary bow shocks, particularly the further the planet is from the Sun. Here, a model beyond the fluid description, namely particle dynamics, must be invoked to compensate for the shortcoming in "heating". This is achieved by means of reflecting a portion of the incoming ions upstream and it can be thought of as the shock decreasing the effective Mach number it is seeing [*Treumann*, 2009]. The reflection is set up by a cross-shock electric field which acts as a potential barrier to the incoming ions. Only the population with sufficient ram energy will be able to overcome this barrier upon the first encounter.

*Ion reflection*

Ion reflection is a fundamental process in supercritical shocks and their associated signatures are prominent in observations [e.g. *Paschmann et al.*, 1981]. For quasi-perpendicular shocks, a fraction of the incoming ions reencounter the shock after their partially-gyrated reflection and are ultimately transmitted downstream with the convected field. These reflections are manifested as an enhancement just preceding the shock ramp. The feature is known as the foot and its locality corresponds to the magnetic field orientation restricting the ions from escaping far enough upstream. The enhancement in the magnetic field arises from the formation of a current layer by the motional electric field. This electric field acts to transversely drift the reflected ions to a velocity $v_y$ [Treumann, 2009]. The current density in this layer is thus $j_y \sim en_{i,refl}v_y$ with the corresponding foot magnitude as $B_z \sim \mu_0 j_y d_x$; where $j_y$ is the

*y*-directed current density, $n_{i,refl}$ is the reflected ion number density, $v_y$ is the drift velocity tangential to the shock plane, and $d_x$ is the shock thickness along the normal. Here the Cartesian coordinate system is such that *x* is along the shock normal, *z* is along the upstream magnetic field and perpendicular to the shock normal, and *y* completes the orthogonal system (note the Cartesian coordinate system for the dataset is defined differently).

Another unique feature is the shock overshoot between the ramp and the downstream field with a lengthscale in the order of an ion gyroradius. One of the earliest detections of the shock overshoot by *Russell and Greenstadt* (1979) highlighted that ion thermalization takes place within the ramp and completes only after the overshoot has finished where irreversibility is eventually imposed. Far downstream from this region and in steady-state, the magnetic field returns to its Rankine-Hugoniot predicted value. The enhancement in the magnetic field at the overshoot can be significantly larger than the Rankine-Hugoniot limit of 4 (for a perfectly perpendicular shock. [e.g. *Bagenal et al.*, (1988); *Masters et al.*, (2013)].

*Shock Geometry*

At distances much greater than the planetary radius ($d \gg R_p$), the shape of a bow shock wave asymptotes to that of a cone formed by the locus of wave fronts with the centre of the planet as the focus point [*Landau and Lifshitz*, 1959]. The angle of this 'Mach Cone', $\mu$, defined as the angle between the edge of the cone and the Sun-planet line, is dependent on the upstream Mach number, *M*, and given by

$$\sin \mu = \frac{1}{M}; M > 1 \qquad (1)$$

This equation is associated with the sonic Mach number, $M_s$, in gas dynamic theory [*Spreiter et al.*, 1966]. In space applications, it has been found to be largely consistent with the fast magnetosonic Mach number, $M_f$, when fitted with observations of distant shock crossings of Venus, Earth and Mars [*Slavin et al.*, 1984]. With increasing heliocentric distance, the Mach cones of each planet were shown to become better in agreement with gas dynamic theory. This is attributed to the decrease in the interplanetary magnetic field (IMF) strength since

$$M_f^2 = \frac{M_s^2 M_A^2}{M_s^2 + M_A^2} \qquad (2)$$

Equation 2 is simply an extension of $v_f^2 = v_a^2 + v_s^2$ and $M = u/v$ where $u$ is the flow speed and $v$ is the characteristic speed. In the limiting case of the IMF strength (and correspondingly the Alfvén speed $v_A$) decreasing, we have $M_f \xrightarrow{B \to 0} M_s$.

At distances comparable to the planetary radius ($R_p < d \lesssim O(10^2)$), a shock wave takes a size and shape similar to the obstacle [*Billig*, 1967]. Saturn's magnetosphere is a blunt body and a detached bow shock is thus formed in the dayside region [*Slavin et al.*, 1985]. Cassini's orbits typically restrict bow shock crossings to the dayside region enabling detailed modelling of its three-dimensional size and shape [*Masters et al.*, 2008; *Went et al.*, 2011; the latter herein referred to as W11].

**Saturn as a unique laboratory for collisionless shocks**

The planets in the solar system are located at heliocentric distances which collectively cover a vast range of solar wind conditions. *In situ* spacecraft observations thus provide an insight into the solar wind properties and their interactions with, for example, planetary magnetospheres. This knowledge can also be extended to exoplanetary systems believed to be similar to those of the solar system. The problem of collisionless shock wave dynamics, especially at high Mach numbers, is of wide interest not only to the solar system community but also to the astrophysicists. Supernova shocks, for example, are characterised by very high Mach numbers and their exploration is only limited to remote observations and simulations. At 10 AU, Saturn is in a unique position in the heliosphere where the Mach numbers are significantly higher than are available at Earth. Saturn's bow shock therefore represents an excellent laboratory for exploring the behaviour of such shocks. *Russell et al.* (1982) showed the typical solar wind dimensionless parameters as a function of heliocentric distance. When the radial profiles of each state variable is examined individually, it can be seen that the Mach number evolution is principally controlled by the wave speeds i.e. $|\Delta v_{f,A,s}(R)| \gg |\Delta u(R)|$. For all Mach numbers, there is a monotonic increase while the $\beta$ changes marginally.

*Scope and limitation of this paper*

Until Cassini, observations of the highest Mach number shocks were made from spacecraft flybys. In this paper, we exploit the long term presence of Cassini in a high Mach number regime of the solar system. We are therefore able to expand on the work by *Achilleos et al*. (2006) which presented a set of several crossings from Saturn Orbital Insertion. This work will characterise Saturn's bow shock using the largest sample of crossings to date and improved techniques from previous works. Statistical studies of Earth's bow shock have been made for modest Mach numbers in the range $M_A$ = 2-8. Here we have a much larger range of Mach numbers spanning two orders of magnitude. A subset of these crossings has been explored in *Sulaiman et al.* (2015) where Saturn's bow shock was discussed as a prototype of high Mach number shocks in the astrophysical-like regime.

In the context of Saturn's magnetosphere, the plasma $\beta$ is expected to be significantly higher in the magnetosheath (region downstream of the bow shock) by virtue of the shock strength [*Schwartz et al.*, 1988]. The $\beta$ condition at the magnetopause has been widely reported to influence the onset of reconnection [e.g. *Swisdak et al.*, 2003]. In addition to the physical uniqueness, Saturn's magnetosheath is geometrically distinct with a non-axisymmetry in the magnetopause shape being manifested in the flow and draping pattern [*Pilkington et al.* (2014); *Sulaiman et al.*, (2014)]. The bow shock and magnetosheath are sites of the external driver and instructive for fully capturing the magnetospheric dynamics via mass, momentum, and energy transfer.

Cassini is a single spacecraft and this comes with its associated limitations. One is the capability to separate spatial and temporal variability in the observations. Direct inference of the shock speed, for example, cannot be made reliably. Multi-spacecraft timings are used at Earth to measure the normal velocity of the bow shock as it propagates between two or more spacecraft. Another limitation is in obtaining particle measurements such as upstream ion temperature and bulk speed; the latter requires the instrument's field-of-view to be in the direction of the incoming flow.

While shock waves can be characterized using several Mach numbers, we will use the Alfvén Mach number, $M_A$, in this paper. Like most studies, $M_A$ is the preferred choice of Mach numbers and the fast magnetosonic speed needed to determine the fast Mach number, $M_f$, is complicated by the non-isotropic nature of MHD modes and thus dependent on the propagation direction upstream, $\theta_{Bn}$. Determination of $M_A$ does not require this parameter nor upstream temperature measurements (a difficulty specific to Cassini, particularly for a large and consistent survey as here).

In this paper, we use magnetic field data [*Dougherty et al.*, 2004] from the years 2004-2014 with over 800 identified shock crossings. These do not include immediately successive crossings so as not to create a bias towards crossings in the same location of parameter space. Figure 1 highlights the spacecraft positions where bow shock crossings were identified relative to Saturn at the origin. A vast majority of the coverage is on the dayside, particularly limited to regions of low to mid-latitudes and roughly equal on dawn and dusk flanks. The magnetic field measurements used are at 1s resolution. We focus our attention on magnetic field signatures of quasi-perpendicular shocks and highlight their evolution with increasing Alfvén Mach numbers.

**Technique**

*Determination of the Shock Normal*

The coordinate system used in this paper is the Cartesian Kronocentric Solar Magnetic (KSM) system [*Dougherty et al.*, 2005] which centres Saturn at the origin, with positive $\hat{X}$ pointing towards the Sun, $\hat{Y}$ orthogonal to the magnetic dipole axis $\hat{M}$ (approximately aligned with the rotation axis at Saturn) and pointing towards dusk, i.e. $\hat{M} \times \hat{X}$, and $\hat{Z}$ chosen such that the magnetic dipole axis $\hat{M}$ is contained in the *X*-*Z* plane with positive $\hat{Z}$ pointing in the northward sense (see Figure 1).

Unlike a normal (or oblique) shock wave, the bow shock's curvature introduces the complexity of global non-uniformity along its surface. In other words, each crossing must be associated with a unique normal vector $\hat{n}$ to the shock's surface at that point (this vector always points upstream). Thus for a given upstream flow vector and IMF direction, the bow shock exhibits a range of $\theta_{Vn}$ and $\theta_{Bn}$ at any given time. This is in a sense analogous to an assembly of many fragments of planar shocks distributed across the surface, each with its specific set of upstream parameters and therefore processing the flow accordingly. The approximation of a planetary bow shock as thin and locally planar is valid since the radius of curvature is much larger than the shock width.

Two common techniques can be employed to determine the local shock normal using a single spacecraft, namely the coplanarity theorem and model boundary equations. The former requires the local magnetic field vectors upstream and downstream to lie in the same plane as the local shock normal [*Abraham-Shrauner*, 1972] (subscripts '*u*' and '*d*' will denote the upstream and downstream regions respectively). This technique, however, breaks down for perfectly parallel and perpendicular shocks, i.e. $\theta_{Bn} = 0°$ and $90°$ respectively. While these two extreme cases are uncommon, uncertainties still prevail. Magnetic field measurements downstream are characterised by large fluctuations and this yields a substantial error associated with measuring $\mathbf{B}_d$. The closeness of the measured $\mathbf{B}_d$ to representing the actual field downstream depends on the interval over which the data is averaged. Doubts in selecting a downstream interval stem from deciding what really is representative of the region "immediately downstream" given that the shock is in continuous motion and whether the selection criterion is consistent throughout all shock crossings. Moreover, for quasi-parallel ($\theta_{Bn} < 45°$) configurations, this difficulty extends to measurements of $\mathbf{B}_u$ where there are large fluctuations ($\delta\mathbf{B}/\mathbf{B}_0 \sim 1$) associated with the foreshock region. This technique is nevertheless broadly used particularly in studies on one or very few shock crossings [e.g. *Achilleos et al.*, 2006].

The second and preferred technique for this work uses the model boundary equation from W11. The equation describing the size and shape of Saturn's bow shock is that of a cylindrically symmetric conic section parameterised by upstream conditions using over 500 crossings. The distance of the subsolar point, $R_{SN}$, was assumed to vary with the upstream dynamic pressure, $P_{dyn}$, according to a power law that was determined empirically as $R_{SN} \propto P_{dyn}^{-1/5.4}$. *Horbury et al.*, (2002) estimated the local normals of 48 quasi-perpendicular shock crossings at Earth using Cluster's fleet of four spacecraft. They compared the normals, determined using inter-spacecraft timings, with each of those estimated using the Coplanarity Theorem and a bow shock model from which the methodology of W11 was inherited [*Formisano*, 1979; *Peredo et al.* 1995]. They found large discrepancies of $22° \pm 4°$ with the Coplanarity Theorem. The comparison with the model, on the other hand, was found to be in remarkably good agreement with ~80% of the sample having a deviation of less than 10°. It is worth noting the Saturnian system undergoes an inherent periodic oscillation with a typical amplitude of 1 and occasionally up to 5 $R_s$ (Saturn equatorial radius = 60,268 km). This has been demonstrated by *Clarke et al.* (2010) and subsequent MHD simulations revealed the bow shock to exhibit this oscillation [e.g. *Jia et al.*, 2012]. The size of the bow shock compared to the variability in the motion, as well as the bluntness of the dayside surface, means the effect of the oscillation on determining the normal geometrically is negligible.

*Determination of the Alfvén Mach Number ($M_A$)*

The Alfvén Mach number, $M_A$, is the key parameter in organising the large sample of shock crossings into regimes in a parameter space. In this way, we are able to focus on a subset as a particular class of shocks and draw comparisons. Recall this quantity is given by

$$M_A \equiv \frac{V}{v_A} = \frac{\sqrt{\mu_0 P_{dyn}}}{B_u} \qquad (3)$$

where $P_{dyn}$ is the upstream ram pressure along the shock normal $\rho V^2 \cos^2\theta_{Vn}$, $\theta_{Vn}$ is the angle between the upstream flow vector $V$, assumed to always be directed along $-X_{KSM}$, and the local normal of the shock surface $\hat{n}$. The ram pressure is estimated from the power law in W11 based on the crossing position of each crossing. Embedded in this relationship are local density measurements and solar wind propagations. By obtaining $P_{dyn}$ and $B_u$ for each crossing, we are then able to calculate $M_A$ using Equation 3.

**Results**

*$M_A$ Parameter Space*

With these estimations, a parameter space can be constructed to show the distribution of $M_A$ (see Figure 2). The median $M_A$ of 14 (red dashed line) is indeed close to the theoretical expectation as inferred from scaling laws of state variables with increasing heliocentric distance (see Figure 1 in *Russell et al.* (1982)). The crossings span across two orders of magnitude of $M_A$ from an Earth-like regime of 2-8 to an astrophysical-like regime of $O(10)$ - $O(10^2)$. It is not only evident from Figure 2 that the typical $M_A$ is higher at Saturn but also that the environment is more variable.

Three quasi-perpendicular shock crossings of different $M_A$ are represented as red markers on the parameter space and the differences in their magnetic field profiles are very obvious in Figure 3 and Table 1. This particular set of crossings (increasing $M_A$ and similar $\theta_{Bn}$) was chosen to compare a pair of similar $B_u$ and a pair of similar $P_{dyn}$. The first is an inbound (i.e. passing from upstream to downstream) crossing of $M_A \sim 5$ and is one typically found in the near-Earth space. It is characterized by a sharp, local transition between both regimes – a feature unique to quasi-perpendicular shocks. As $M_A$ increases in the second and third panels, ~22 and ~38 respectively, there is an increasing trend in the downstream variability, maximum field and prominence of the foot region preceding the ramp. Here, ion dynamics becomes important and this is discussed in more detail in *Sulaiman et al.* (2015).

Figure 4 (a-d) summarises the sample of crossings into four distributions of $\theta_{Bn}$, $B_u$, standoff distance $R_{bs}$ and $M_A$. Figure 4a reveals that the quasi-perpendicular configuration is the most observed at Saturn's bow shock with 9% for $\theta_{Bn} < 45°$, 81% for $\theta_{Bn} \geq 45°$ and 50% for $\theta_{Bn} \geq 70°$. This is attributed to the Parker spiral at 10 AU being significantly more azimuthal. Another factor is the orbit of Cassini which crosses the bow shock mostly in the dayside where the surface is blunt. For these reasons, we expect the detection of a foreshock region to be uncommon though this has been reported nonetheless [*Bertucci et al.*, 2007; *Andrés et al.*, 2013]. Another possibility for this could be from the fluctuations in the IMF. We expect the favourable IMF direction to remain steady long enough for a foreshock to be set up and detected far upstream by Cassini. This is beyond the scope of this paper; however a study of δ**B**(t) upstream of Saturn is a topic of interest for future study.

Recalling that the standoff distance is used as a proxy for the upstream ram pressure, the distribution in $M_A$ is principally controlled by the behavior of $B_u$. This is because the spread in $B_u$ spans two orders of magnitude upstream of Saturn's bow shock compared with $\rho u^2$ varying by only one order of magnitude. Another reason is because the $M_A$ changes as $1/B_u$ and only $\sqrt{\rho u}$. The largest Mach numbers, in particular, are much more likely to be attributed to very low IMF strengths rather than unusually large ram pressures (e.g. arrival of ICMEs). For example, a typical $B_u$ of 4 nT at Earth requires a change of 2 nT for a 50% change in $M_A$, whereas at Saturn a typical $B_u$ of 0.6 nT requires a change of 0.3 nT for the same change in $M_A$.

*Shock Relative Overshoot and Varialibty*

It has been established that for super-critical quasi-perpendicular collisionless shocks, microphysics must be invoked to account for the deficit in dissipation that cannot be accommodated by the hydrodynamic formalism i.e. the Rankine-Hugoniot equations [*Scudder et al.*, 1986]. The shock overshoot is a unique and important feature of such shocks immediately succeeding the ramp with a length scale in the order of an ion gyroradius. One of the earliest detections of the shock overshoot by *Russell and Greenstadt* (1979) highlighted that ion thermalization takes place within the ramp and completes only after the overshoot has finished. Far downstream from this region and in steady-state, the magnetic field returns to its Rankine-Hugoniot predicted value.

The overshoot for each crossing can be quantified in a number of ways and here we choose to be consistent with the widely used Relative Overshoot Amplitude (ROA) which is given by

$$ROA = \frac{B_{max} - \langle B_d \rangle}{\langle B_d \rangle} \qquad (4)$$

where $B_{max}$ is the highest field strength recorded in the crossing and $B_d$ is the downstream field. This is averaged over an interval far enough from the overshoot-undershoot region but close enough to the shock ramp to give a better representation of the downstream magnetic field strength.

Figure 5a presents the ROA calculations for quasi-perpendicular shocks as a function of $M_A$. Table 1 provides a summary of the corresponding statistical measures. It is clear at first from Figure 5a that the super-criticality of Saturn's bow shock prevails with all crossings having a positive non-zero ROA. The positive correlation here is moderate-to-strong and the unprecedented range of $M_A$ and sample size may confirm that the Mach number is indeed the principal controller of the overshoot. The view here leans much more towards the $M_A$ dependence argument, in spite of the absence of $\beta$ measurements. While $\beta$ and $M_A$ are interlinked and we expect on average the high (low) $M_A$ to be associated with higher (lower) $\beta$, the range of $\beta$ is significantly shorter than that of $M_A$ in the solar wind near Saturn [*Jackman and Arridge*, 2006]. More convincingly from a theoretical perspective, the Mach number takes into account the speed of the flow which in a sense is a measure of the mass flux that requires dissipation across the shock. The Mach number can be arbitrary and hence there is no upper limit on how much mass flux enters the shock to subsequently be dissipated by anomalous resistivity and particle trajectory; ensuring the adequately heated and decelerated flux downstream for a sub-fast magnetosonic regime. Since the Rankine-Hugoniot equations dictate a maximum field enhancement downstream of 4 (assuming $\gamma = 5/3$ for a perpendicular shock) for the conservations law to be maintained, any increase above this limit must be attributed to an additional processes beyond the classical fluid framework. As a result, increases in the Mach number must intensify the role of such additional processes, one of which is manifested as the overshoot.

Figure 5c, on the other hand, shows no clear correlation between ROA and $\theta_{Bn}$ for quasi-perpendicular shocks. This is broadly consistent with the simulation runs by *Tiu et al.* (2011) where they concluded that the overshoot is insensitive to $\theta_{Bn}$. $\theta_{Bn}$ is a geometric rather than a physical factor like the Mach number or $\beta$. In the context of its role as a control parameter, changes in this quantity (assuming all other parameters are held fixed), affect only the pathway the ions take to achieve the required thermalization.

The variability downstream is also quantified using the root-mean-square (RMS) value of the same downstream interval normalised by the average of the upstream and downstream magnetic field magnitudes (away from the foot and overshoot-undershoot regions). This is given by

$$B_{ave} = \frac{1}{2}(\langle B_u \rangle + \langle B_d \rangle) \qquad (5)$$

Equation 5 corrects for the fact that two shocks can have the same $M_A$ and $\theta_{bn}$ but different $B_u$ and transition profiles thus allowing for better comparison of underlying trends with the control parameters.

Figure 5b presents the normalised downstream RMS as a function of $M_A$ for the same quasi-perpendicular shocks. It is clear that with increasing $M_A$ the variability in the signal becomes more pronounced. The variability is interpreted as spatio-temporal substructures which can be used as an indicator of a shock's departure from a one-dimensional surface. What is observed could possibly be a combination of scaled length effects [*Bale et al.*, 2003; *Scholer and Burgess*, 2006], rates of change of the structure and time spent within the shock layer [*Burgess*, 2006]. We expect as a result some observational bias in the data, for example low vs high speed shock crossings.

Again, Figure 5d reveals no obvious correlation between the normalised downstream RMS and $\theta_{Bn}$. While the full range of $\theta_{Bn}$ is not shown, the variability is expected to be more pronounced when $\theta_{Bn}$ is less than 45°. This is due to the generation of wave structures in a quasi-parallel regime. As for the quasi-perpendicular regime, the argument for the lack of a visible trend in Figure 5d is probably the same as stated for Figure 5c.

**Conclusions**

- We have accumulated a very large sample of Saturn's bow shock crossings and have presented distributions of the upstream conditions. Most of the crossings were in a quasi-perpendicular configuration by virtue of a combination of the Parker spiral at 10 AU and the location of the crossings taking place on the dayside of Saturn.

- Using the upstream magnetic field strength and estimated dynamic pressure of each crossing, the Alfvén Mach number was obtained. This technique, presented here for the first time, has the advantage of overcoming the difficulty and limitations of Cassini's plasma instrument to obtain a Mach number for each of the 871 shock crossings, as presented in Figure 2. This has enabled a study of the overall characteristic of Saturn's bow shock and, more interestingly, laid the foundation for a natural follow-up detailed study of the highest Mach number regime as presented in *Sulaiman et al.* (2015).

- The Alfvén Mach numbers determined show a median value significantly higher than at Earth ($M_A = 14$) and the largest range seen at any planet, owing to the upstream magnetic field.

- The magnetic overshoot was shown to correlate quite strongly with $M_A$ across the entire range and we conclude that it is most likely the primary controller. Although a $\beta$ dependence has been reported, the absence of the parameter in this study is not likely to make it any less instructive since its range is similar at all planets [*Russell et al.*, 1982].

The author is grateful to J. P. Eastwood and J. A. Slavin for useful discussions. We acknowledge the support of MAG data processing/distribution staff. Cassini magnetometer data are publicly available via NASA's Planetary Data System. This work was supported by UK STFC and research grants to Imperial College London as well as travel support from the Royal Astronomical Society.

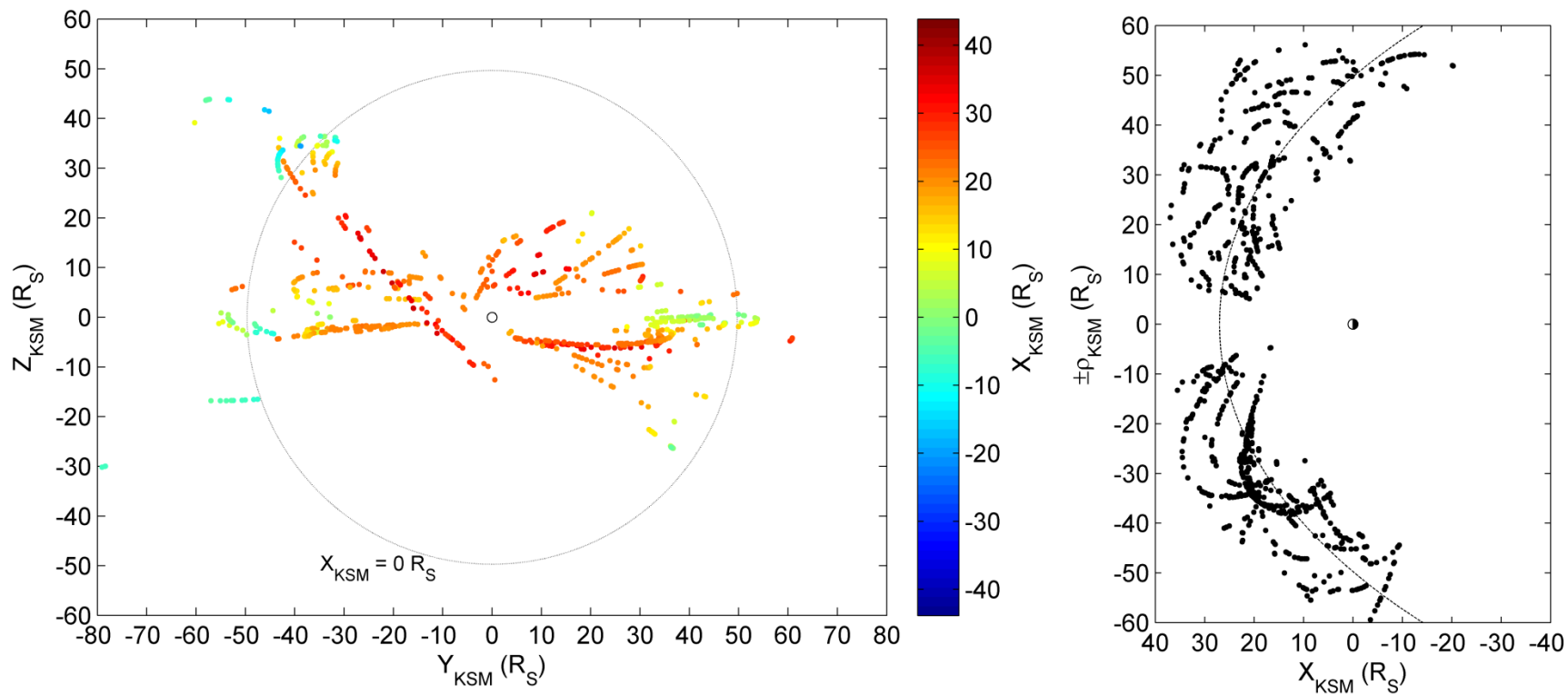

Figure 1.: An overview of the positions of observed bow shock crossings between the years 2004 and 2014 inclusive. These are projected onto a) the *Y-Z* and b) the *X-ρ* planes, where $\rho = \sqrt{Y^2 + Z^2}$. In both figures, the projections of the *Went et al.* (2011) bow shock model is shown with a subsolar distance of 27 $R_s$ corresponding to a solar wind dynamic pressure of ~0.04 nPa.

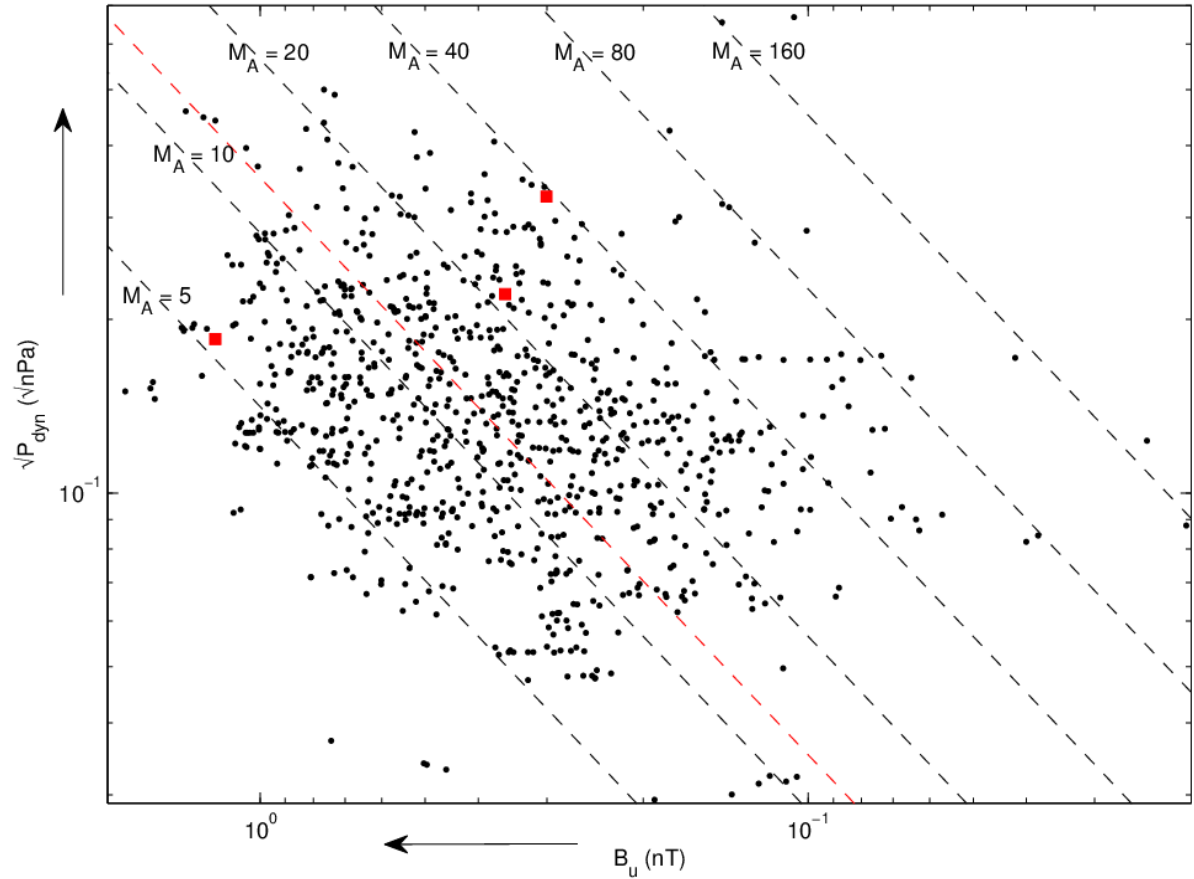

Figure 2.: $M_A$ parameter space of $\sqrt{P_{dyn}}$ vs $B_u$ on a log-log scale. Note abscissa is increasing from right to left. Each marker represents a shock crossing. The three red markers are examples in Figure 3. Contours are overlaid as black dashed lines satisfying Equation 3 and represent lines of constant $M_A$. The red dashed line is the median $M_A$ of 14.

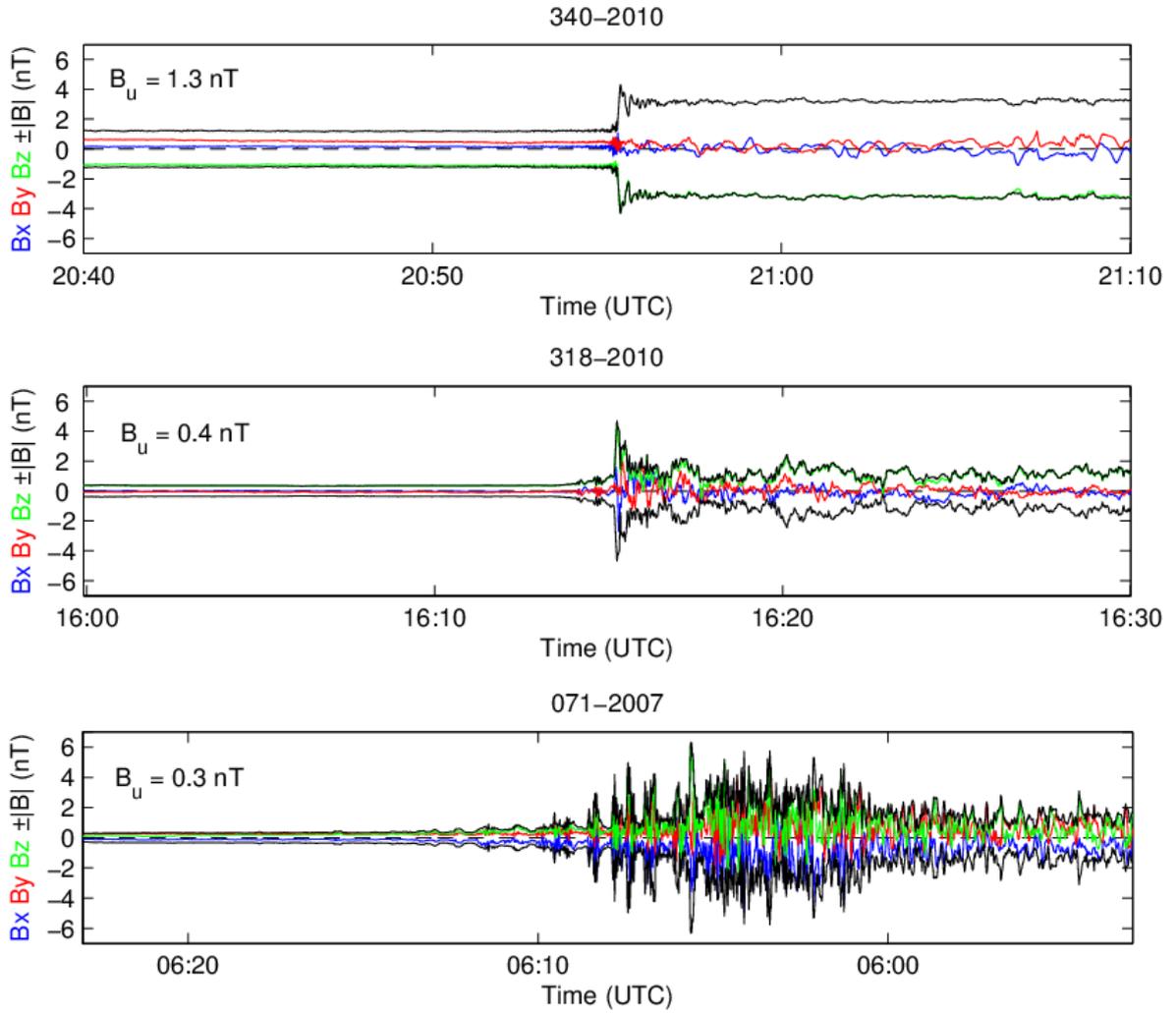

5
6
7
8
9
10

Figure 3.: Three magnetic field time series of quasi-perpendicular shock crossings of increasing $M_A$ corresponding to the three red markers on Figure 4.6. From top to bottom: $M_A \sim 5$ & $\theta_{Bn} = 65°$, $M_A \sim 22$ & $\theta_{Bn} = 81°$ and $M_A \sim 33$ & $\theta_{Bn} = 77°$. The top two panels are inbound crossings. The bottom panel is an outbound panel with the time series reversed.

11
12
13
14

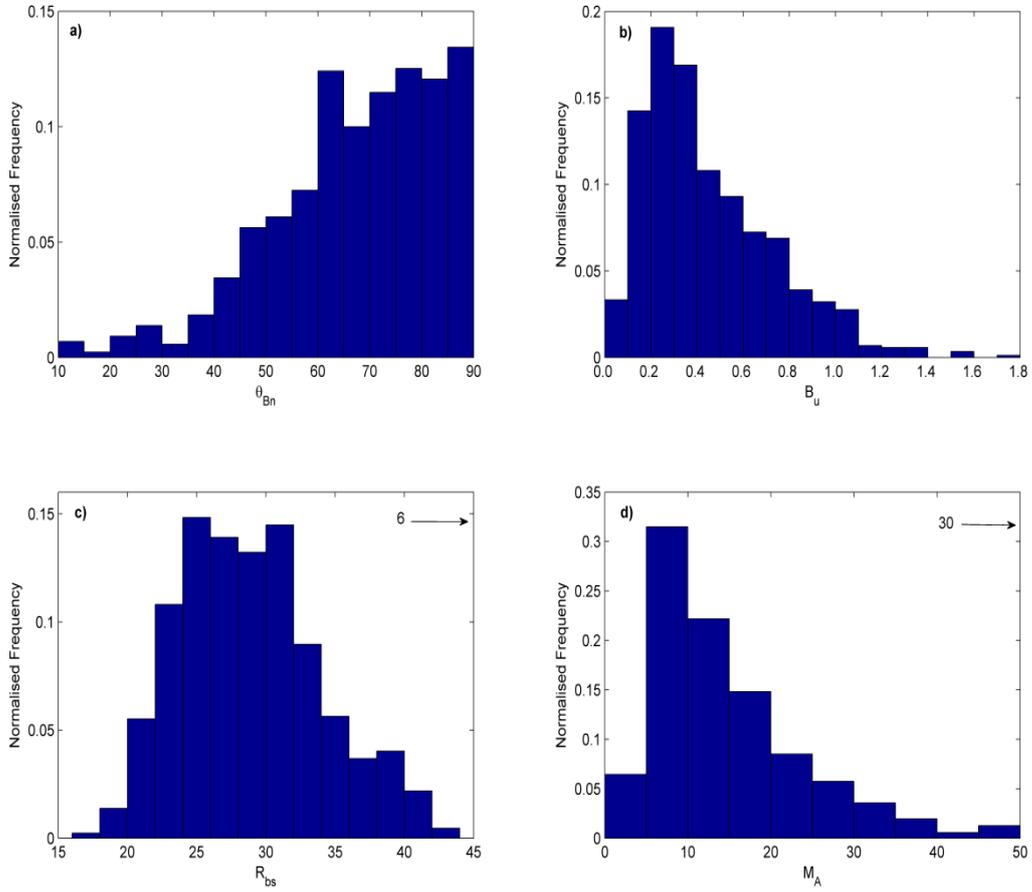

Figure 4.: Normalised frequency distributions of all (871) crossings of Saturn's bow shock for a) $\theta_{Bn}$ with bin width of 5° b) $B_u$ with bin width of 0.1 nT c) standoff distance $R_{bs}$ with bin width of 2 $R_S$ and d) $M_A$ with bin width of 5.

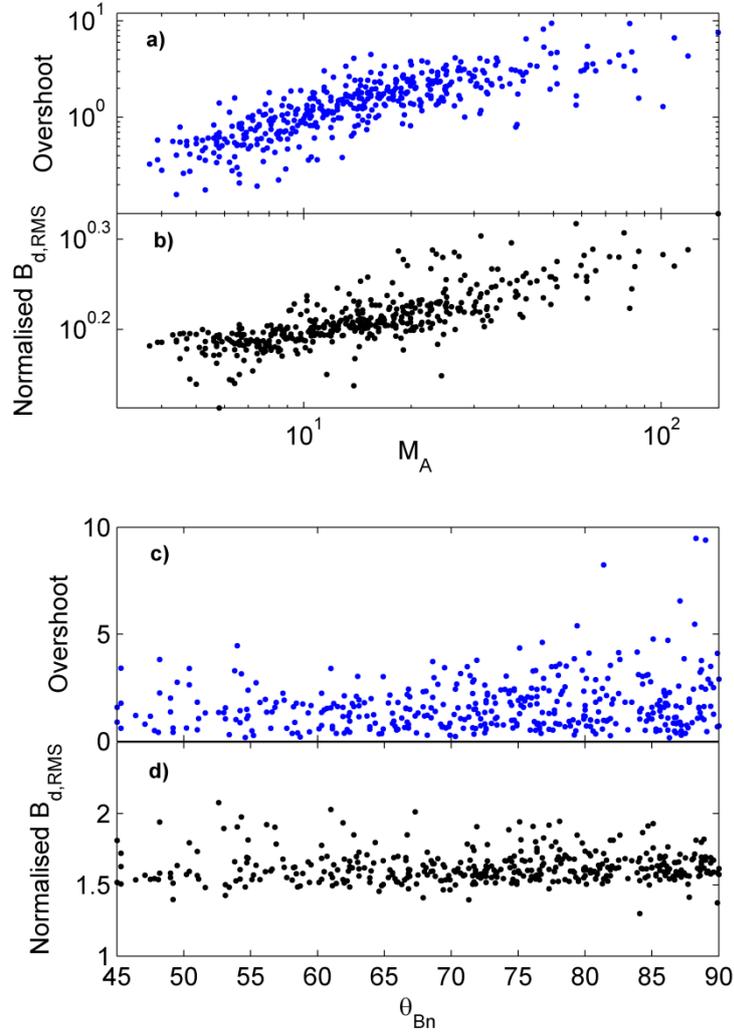

Figure 5.: (a) Overshoot and (b) variability plotted against $M_A$ for highly quasi-perpendicular shocks $\theta_{Bn} \geq 70$. (c) Overshoot and (d) variability plotted against $\theta_{Bn}$.

| | 1 | 2 | 3 |
|---|---|---|---|
| $|B_u|$ (nT) | 1.3 | 0.4 | 0.3 |
| $|B_d|$ (nT) | 3.2 | 1.1 | 1.3 |
| $B_{max}$ (nT) | 4.3 | 4.7 | 6.3 |
| $B_u$ (nT) | [0.16, 0.55, -1.06] | [-0.016, -0.07, 0.35] | [-0.1, 0.24, 0.15] |
| $\hat{n}$ | [0.69, 0.72, -0.01] | [0.73, 0.68, 0.017] | [0.84, 0.41, 0.355] |
| $\theta_{Bn}$ (deg) | 65 | 81 | 77 |
| $M_A$ | 5 | 22 | 33 |

The vectors are in the KSM cartersian coordinate system defined in the text. $\theta_{Bn}$ is the acute angle between the *normalized* $B_u$ and $\hat{n}$.

Table 1: Properties of the three crossings in Figure 3.